\documentclass[aps,twocolumn,prl,superscriptaddress,amssymb,showpacs]{revtex4-1}
\usepackage{graphicx}
\usepackage{amsmath}
\usepackage{comment}
\usepackage{bm}
\usepackage{color}
\usepackage{txfonts}

\begin{document}

\title{High-field Magnetism of the Honeycomb-lattice Antiferromagnet Cu$_2$(pymca)$_3$(ClO$_4$)}

\author{Akira Okutani} 
\affiliation{Center for Advanced High Magnetic Field Science (AHMF), Graduate School of Science, Osaka University, 1-1 Machikaneyama, Toyonaka, Osaka 560-0043}

\author{Takanori Kida} 
\affiliation{Center for Advanced High Magnetic Field Science (AHMF), Graduate School of Science, Osaka University, 1-1 Machikaneyama, Toyonaka, Osaka 560-0043}

\author{Yasuo Narumi} 
\affiliation{Center for Advanced High Magnetic Field Science (AHMF), Graduate School of Science, Osaka University, 1-1 Machikaneyama, Toyonaka, Osaka 560-0043}

\author{Tokuro Shimokawa} 
\affiliation{Okinawa Institute of Science and Technology Graduate University, Onna, Okinawa 904-0395}

\author{Zentaro Honda} 
\affiliation{Graduate School of Science and Engineering, Saitama University, 255 Shimo-Okubo, Sakura-ku, Saitama 338-8570}

\author{Koichi Kindo} 
\affiliation{Institute for Solid State Physics, The University of Tokyo, 5-1-5 Kashiwano-ha, Kashiwa, Chiba 277-8581}

\author{Takehito Nakano} 
\affiliation{Graduate School of Science, Osaka University, 1-1 Machikaneyama, Toyonaka, Osaka 560-0043}

\author{Yasuo Nozue} 
\affiliation{Graduate School of Science, Osaka University, 1-1 Machikaneyama, Toyonaka, Osaka 560-0043}

\author{Masayuki Hagiwara} 
\email{hagiwara@ahmf.sci.osaka-u.ac.jp}
\affiliation{Center for Advanced High Magnetic Field Science (AHMF), Graduate School of Science, Osaka University, 1-1 Machikaneyama, Toyonaka, Osaka 560-0043}


\begin{abstract}
We report on the experimental results of magnetic susceptibility, specific heat, electron spin resonance (ESR), and high-field magnetization measurements on a polycrystalline sample of the spin-$1/2$ distorted honeycomb-lattice antiferromagnet Cu$_2$(pymca)$_3$(ClO$_4$). Magnetic susceptibility shows a broad peak at about 25~K, which is typical of a low dimensional antiferromagnet, and no long range magnetic order is observed down to 0.6~K in the specific heat measurements. Magnetization curve up to 70~T at 1.4~K shows triple stepwise jumps. Assuming three different exchange bonds $J_{\rm A}$, $J_{\rm B}$ and $J_{\rm C}$ from the structure, the calculated magnetization curve reproduces the observed one when $J_{\rm A}/k_{\rm B} = 43.7~{\rm K}$, $J_{\rm B}/J_{\rm A} = 1$ and $J_{\rm C}/J_{\rm A} = 0.2$ except the magnetization near 70~T, where the observed magnetization indicates another step while the calculated magnetization becomes saturated. The relationship between magnetization plateaus and exchange bonds is discussed based on the numerical calculations.
\end{abstract}%
\maketitle


A honeycomb-lattice antiferromagnet (HLA) with the nearest neighbor (NN) exchange interactions is a bipartite antiferromagnet, and thus no magnetic frustration occurs. However, the HLA has the minimum coordination number $z=3$ of exchange bonds among two dimensional (2D) antiferromagnets, indicating quantum fluctuations must be larger than the other bipartite 2D antiferromagnets, such as a square-lattice antiferromagnet. When the distant antiferromagnetic exchange interactions such as the next nearest neighbor (NNN) and the third nearest neighbor interactions are included, the HLA system is expected to exhibit geometrical frustration. A recent theory predicted a rich phase diagram depending on the magnitudes of distant interactions \cite{PRB86_144404}. The ground state of the HLA with different NN and identical NNN exchange interactions was also investigated theoretically \cite{PRB74_140402R}. The appearance of disordered phases and spin liquid states are caused by the geometrical frustration and distortion. \par
In contrast to extensive theoretical studies on HLAs, a few inorganic honeycomb-lattice compounds have been reported to date. For example, in In$_3$Cu$_2$VO$_9$, the Cu$^{2+}$($S=1/2$) ions form a honeycomb lattice that is well separated by InO$_6$ and VO$_5$ layers, thereby providing good two dimensinality \cite{JMMM290-291_310,4}, but this compound exhibits a conventional N\'eel-type collinear antiferromagnetic long range order below 20~K \cite{5}. As the second case, Na$_3$Cu$_2$SbO$_6$ possesses a honeycomb lattice of Cu$^{2+}$($S=1/2$) ions, but its magnetic properties are interpreted as the spin-gapped antiferromagnetic-ferromagnetic bond alternating chain \cite{6}. Another example is Bi$_3$Mn$_4$O$_{12}$(NO$_3$), in which the Mn$^{4+}$($S=3/2$) ions possess a disordered ground state at low temperatures \cite{7} and shows an unusual field-induced long range order \cite{8}. These varieties of magnetic properties in honeycomb-lattice compounds are driven from not only distant exchange interactions but also irregularity of NN exchange interactions. \par
Recently, a honeycomb lattice with bond-dependent exchange interactions $J_{\rm x}$, $J_{\rm y}$ and $J_{\rm z}$ has been extensively studied as a Kitaev model. The important point is that the Kitaev model has an exactly solvable spin liquid ground state \cite{9}, and gapless or gapped Majorana fermion excitations are expected depending on the parameters under the condition of $J_{\rm x}+J_{\rm y}+J_{\rm z}=1$. After a theoretical study by Jackeli and Khaliullin \cite{10}, this Kitaev model has been expected to be realized in some Ir oxides and $\alpha$-RuCl$_3$ \cite{11,12}. \par
The honeycomb-lattice compound Cu$_2$(pymca)$_3$(ClO$_4$) (pymca: pyrimidine-2-carboxylate) was reported to be a regular honeycomb-lattice copper compound which crystallizes in a trigonal crystal system with space group $P31m$ and shows no magnetic long range order down to 2~K, although distant exchange interactions are not expected from the crystal structure \cite{13}. Therefore, this compound was expected to be the second example of a quantum spin-orbital liquid material composed of Cu ions followed by 6H-Ba$_3$CuSb$_2$O$_9$ \cite{14,15,16,17,18}. However, the latest structural report \cite{19} provided by synchrotron X-ray diffraction data invalidated this structure and clarified it as a distorted honeycomb lattice in this compound. More specifically, there are two crystallographically inequivalent Cu sites (Cu1 and Cu2) in this compound, and the atomic distances between Cu1 and Cu2 ions are $5.5932(11)~\AA$, $5.5112(12)~\AA$, and $5.5083(15)~\AA$ \cite{19}. It is found that the latter two Cu-Cu distances are close to each other. \par
In this letter, we report on the experimental results of magnetic susceptibility, specific heat, electron spin resonance (ESR), and high field magnetization measurements of polycrystalline samples of the $S$=1/2 distorted HLA Cu$_2$(pymca)$_3$(ClO$_4$). First, all these experimental methods and analytical method of the magnetization will be presented. Next, the experimental results and the analysis of the magnetization will be provided.  After the comparison between the observed magnetization and its numerical analysis, we will discuss the relationship between the observed magnetization plateaus, one-third and two thirds plateaus of the saturation magnetization,  and the exchange bonds based on the numerical calculations. The final paragraph will be devoted to the conclusions of this work.


\begin{figure}
\includegraphics[width=\linewidth]{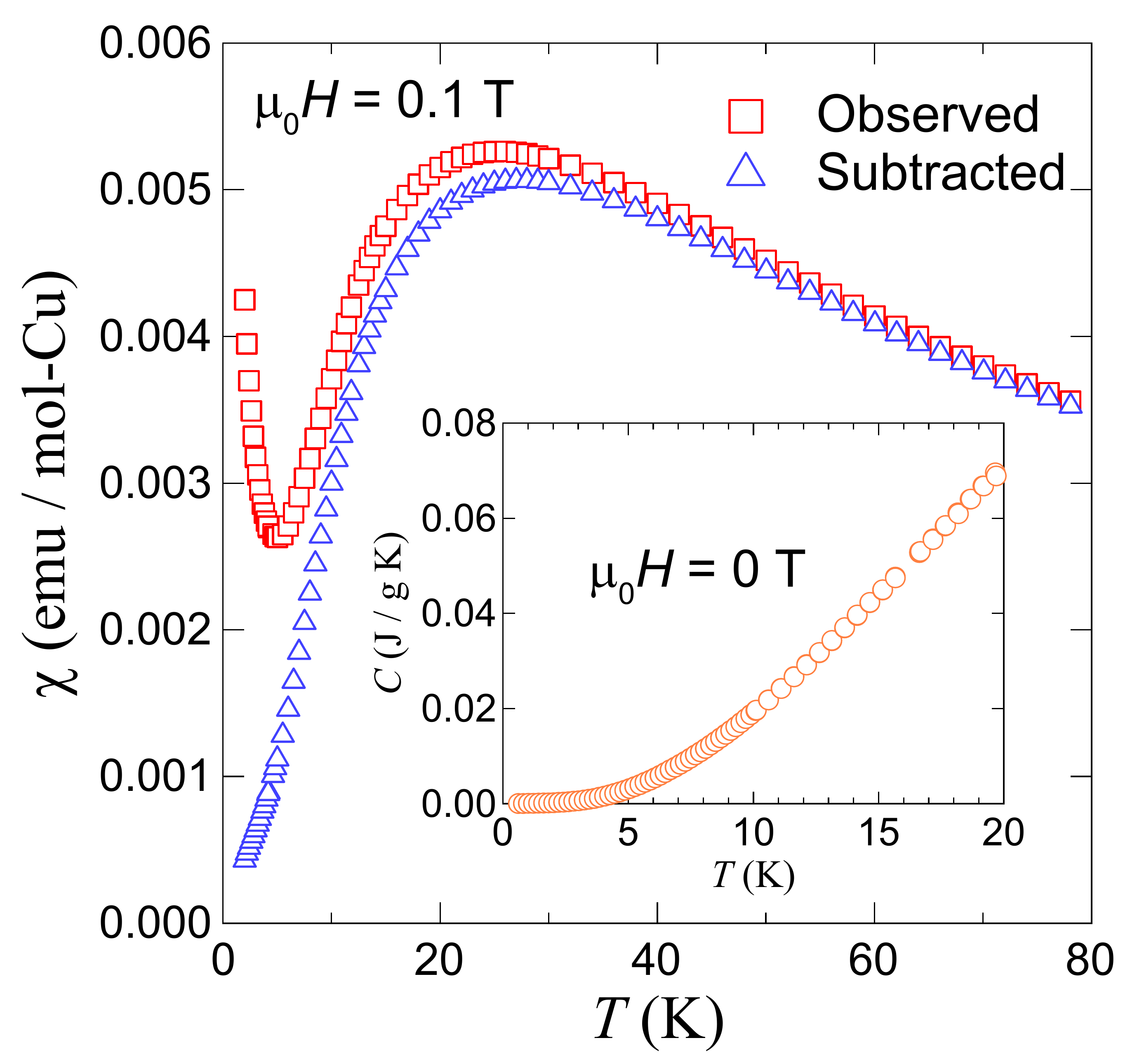}
\caption{Temperature dependences of magnetic susceptibility ($\chi\left(=M/H\right)$) and specific heat ($C$) at zero field (inset) of a polycrystalline sample of Cu$_2$(pymca)$_3$(ClO$_4$). Open squares indicate observed magnetic susceptibility data and open triangles indicate the magnetic susceptibility data subtracted magnetic impurity component by the Curie term from the observed data.}
\label{f1}
\end{figure}

Polycrystalline samples of Cu$_2$(pymca)$_3$(ClO$_4$) were synthesized by the hydrothermal reaction according to the method reported in Ref. \cite{13}, and the samples were characterized by the XRD technique. Magnetic susceptibility measurements were performed at the temperature from 1.9~K to 300~K using a superconducting quantum interference device (SQUID) magnetometer (Quantum Design MPMS XL-7). Specific heat measurements were conducted using PPMS measurement system by relaxation method at the temperatures down to 0.6~K. \par
ESR measurements were done by using a X-band ESR spectrometer (Bruker EMX) together with a He-flow cryostat (Oxford Instruments). High-field magnetization measurements were carried out with an induction method using a pick-up coil at 1.4~K and 4.2~K in pulsed magnetic fields of up to nearly 70~T at AHMF in Osaka University. Observed magnetization curves were analyzed by a Quantum Monte Carlo (QMC) method using ALPS package \cite{20}. Our QMC code is based on the directed loop algorithm in the stochastic series expansion representation \cite{21,22,23}. The calculation was performed for a system size of 72 under the periodic boundary condition.


Figure \ref{f1} shows the temperature dependence of magnetic susceptibility ($\chi=M/H$ where $M$ is the magnetization and $H$ is the external magnetic field) of a polycrystalline sample of Cu$_2$(pymca)$_3$(ClO$_4$) (open squares) measured at $\mu_{0}H = 0.1~{\rm T}$. The magnetic susceptibility has a broad maximum near 25~K, which is typical of a low dimensional antiferromagnet, and a steep increase below 5~K, which might be the susceptibility from a paramagnetic impurity. To obtain the intrinsic magnetic susceptibility of the sample, we subtracted the paramagnetic impurity component by the Curie term from the observed magnetic susceptibility (open triangles). Here, we assume the paramagnetic component with $S=1/2$ which is expressed by $\alpha C/T$, where $C$ is the Curie constant, with $g$-value $g = 2.13$ (from multi-frequency ESR measurements at the lowest temperatures, 3.0~K for X-band and 1.5~K for other frequencies), and impurity concentration $\alpha = 1.8~\%$. As a result, the intrinsic magnetic susceptibility shows a monotonical decrease toward zero susceptibility upon cooling from 20~K. From the Curie-Weiss fitting of the inverse susceptibility between 100 and 300~K, we obtained the Weiss temperature ${\it \Theta} = -49.3~{\rm K}$, indicating that antiferromagnetic interactions are dominant in this compound. The inset of Fig. \ref{f1} indicates the temperature dependence of the specific heat (open circles) measured at zero field. No anomalies and peaks were observed and thus there is no evidence for a long range magnetic order. \par

\begin{figure}
\includegraphics[width=\linewidth]{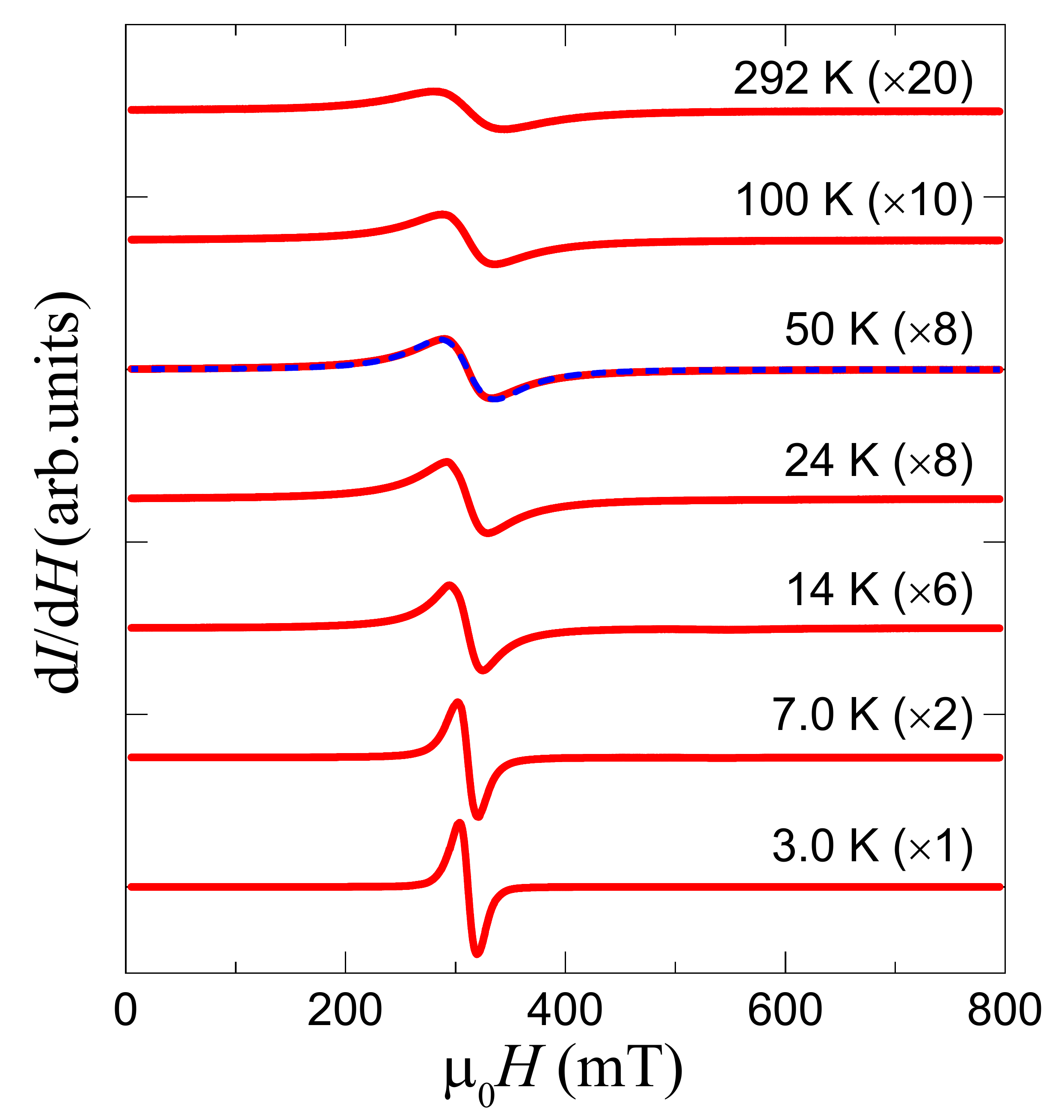}
\caption{X-band ESR spectra of a polycrystalline sample of Cu$_2$(pymca)$_3$(ClO$_4$) measured at the designated temperatures. The broken line at 50~K indicates the line fitted with a single Lorentzian function. The numbers with cross inside the parentheses next to the temperatures are scaling factors to make the spectra easy to see.}
\label{f2}
\end{figure}

X-band ESR spectra of a polycrystalline sample of Cu$_2$(pymca)$_3$(ClO$_4$) at the designated temperatures are shown in Fig. \ref{f2}. All the ESR spectra can be fitted with a single Lorentzian function, indicating an exchange coupled isotropic system, and we can get the $g$-value from the fitting. We have found that the $g$-value (2.13) is almost temperature independent between 3 and 292~K. \par

\begin{figure}
\includegraphics[width=\linewidth]{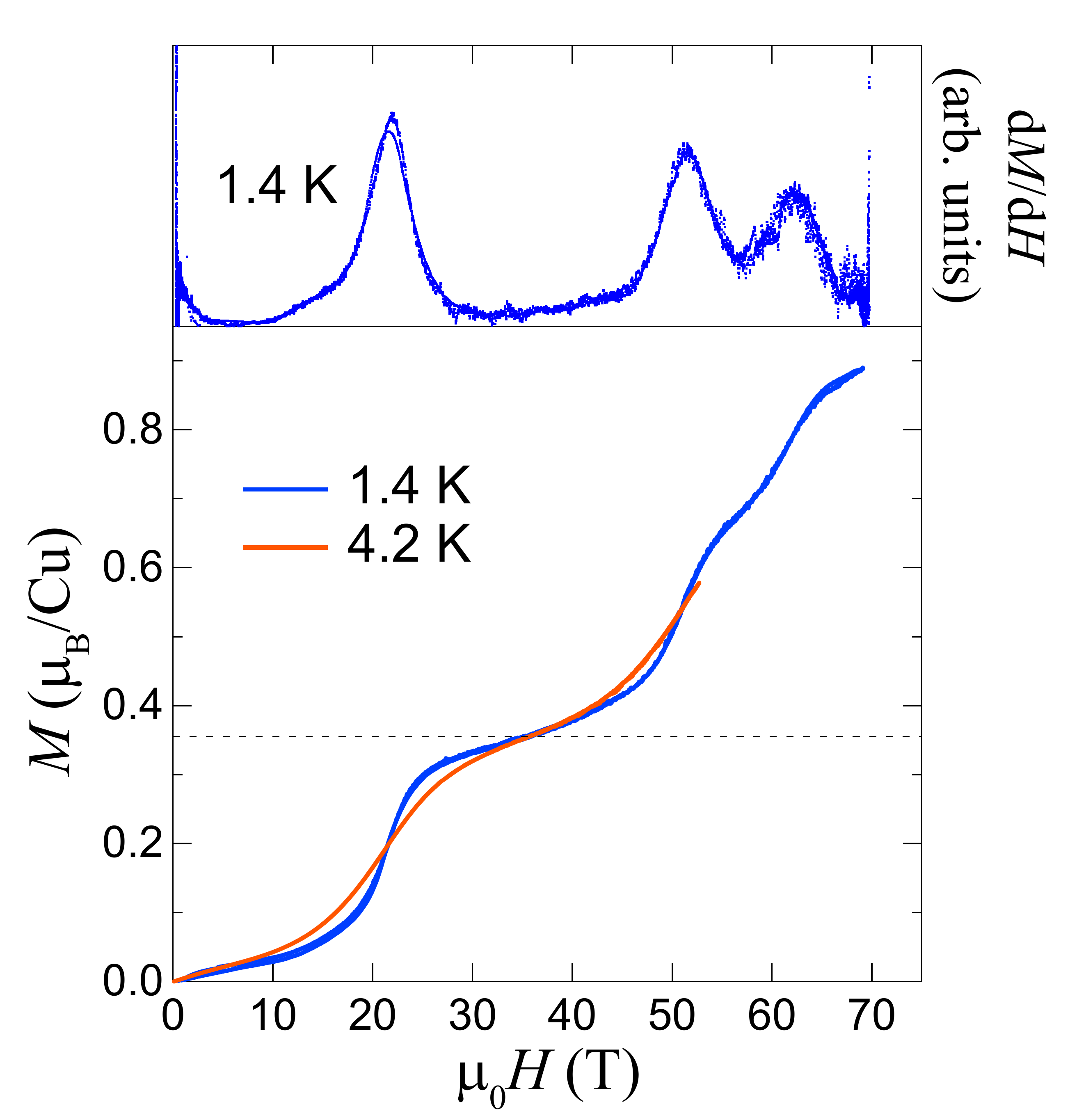}
\caption{High field magnetization ($M$) curves of a polycrystalline sample of Cu$_2$(pymca)$_3$(ClO$_4$) in high magnetic fields of up to 70~T at 1.4~K and up to 50~T at 4.2~K (lower panel). The broken line is one-third of the saturation magnetization expected from the $g$-value. The field derivative of magnetization curve (${\rm d}M/{\rm d}H$) at 1.4~K is also indicated in the upper panel. The three peaks corresponding to stepwise increases in magnetization curves are clearly observed.}
\label{f3}
\end{figure}

Figure \ref{f3} depicts magnetization curves of Cu$_2$(pymca)$_3$(ClO$_4$) at 1.4~K and 4.2~K (lower panel) and the field derivative of the magnetization at 1.4~K (upper panel) measured in pulsed magnetic fields. The triple stepwise increases are observed in the magnetization curve at 1.4~K, and correspondingly three peaks in the ${\rm d}M/{\rm d}H$ curve at 1.4~K are clearly observed. The magnetization values at the flat stages at about 30~T and 60~T are one third and two thirds of the saturation value evaluated from the $g$-value (2.13) of this compound. The magnetization value at about 70 T is not the saturation value ($1.07\mu_{\rm B}/{\rm Cu}^{2+}$). This magnetization curve is not explained by the magnetization of the $S=1/2$ simple HLA with uniform exchange interactions, which will be calculated and shown as a monotonical increase up to the saturation field in the next paragraph. \par

\begin{figure}
\includegraphics[width=\linewidth]{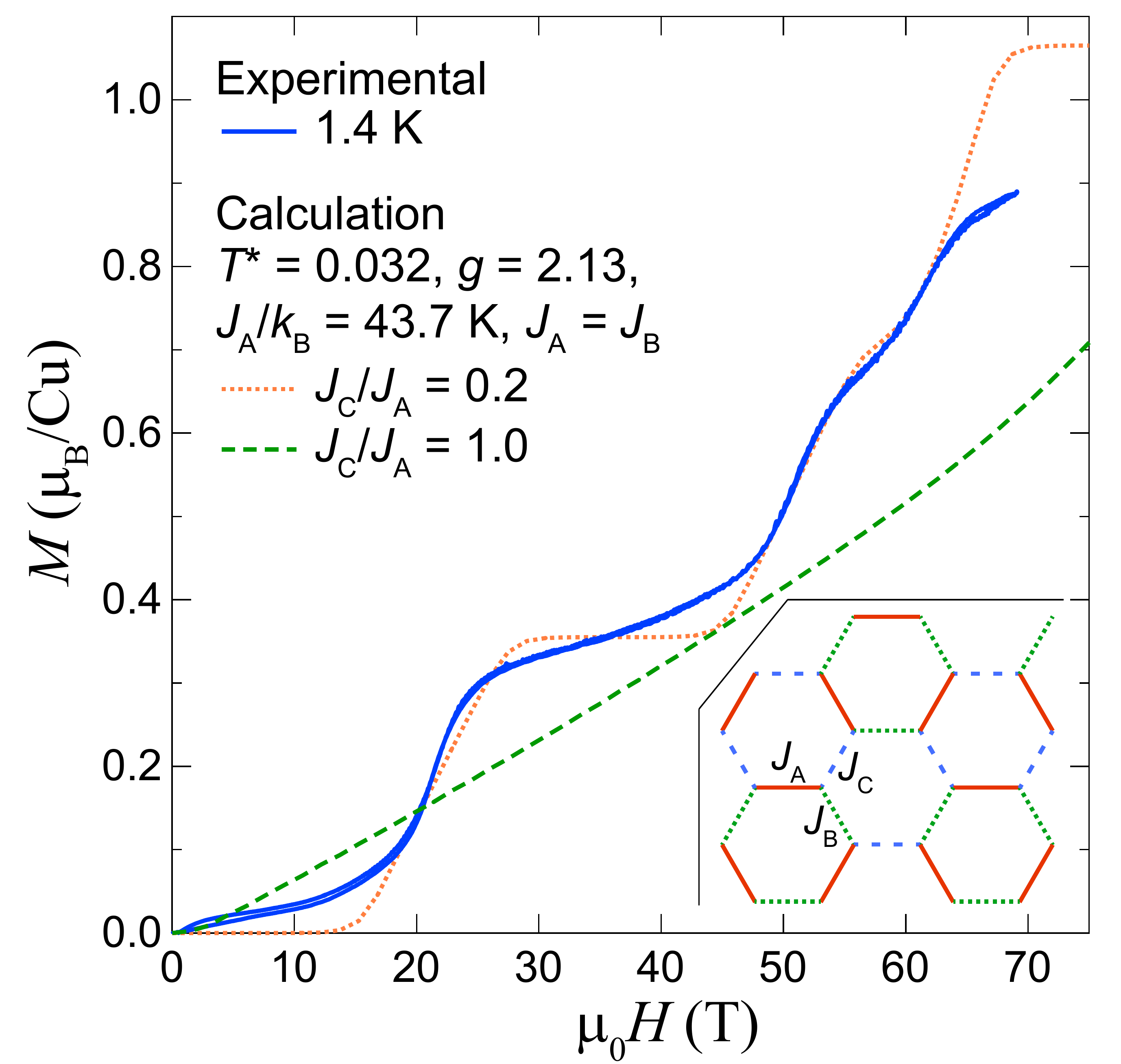}
\caption{High field magnetization curve of a polycrystalline sample of Cu$_2$(pymca)$_3$(ClO$_4$) at 1.4~K (solid line) and those of a honeycomb lattice antiferromagnet calculated by a Quantum Monte Carlo method using ALPS package \cite{20} for different sets of exchange constants given in the panel (dotted and broken  lines).}
\label{f4}
\end{figure}

Next, we calculate the magnetization curves by means of a QMC method \cite{21,22,23} using ALPS package \cite{20}. Here, we assume three different exchange constants $J_{\rm A}$, $J_{\rm B}$ and $J_{\rm C}$ as a HLA model as shown in the inset of Fig. \ref{f4}. From the structural analysis, three different exchange constants are expected, but two of them must be nearly identical because of the local environment around Cu ions and the bonding between Cu ions. At the beginning, we calculate the magnetization curve for a regular HLA in which the magnitudes of all the exchange interactions are identical, $J_{\rm A} = J_{\rm B} = J_{\rm C}$, as shown by the broken line in Fig. \ref{f4}. This magnetization curve continuously increases from zero with a concave curvature, which is caused by quantum fluctuations in a low dimensional antiferromagnet, with increasing magnetic fields.  From the experimental results of magnetic susceptibility, specific heat and magnetization measurements, the ground state must be a nonmagnetic singlet state with an excitation energy gap to the triplet state. In order to open the energy gap, different exchange interactions should be included in the calculation. The calculated magnetization curve reproduces the observed magnetization curve when $J_{\rm A}/k_{\rm B} = 43.7~{\rm K}$, $J_{\rm A} = J_{\rm B}$ and $J_{\rm C}/J_{\rm A} = 0.2$.  However, above the $2/3$ magnetization plateau, the calculated magnetization is completely different from the observed one. The calculated curve increases toward the saturation value, while the observed one shows another magnetization plateau. \par

\begin{figure}
\includegraphics[width=\linewidth]{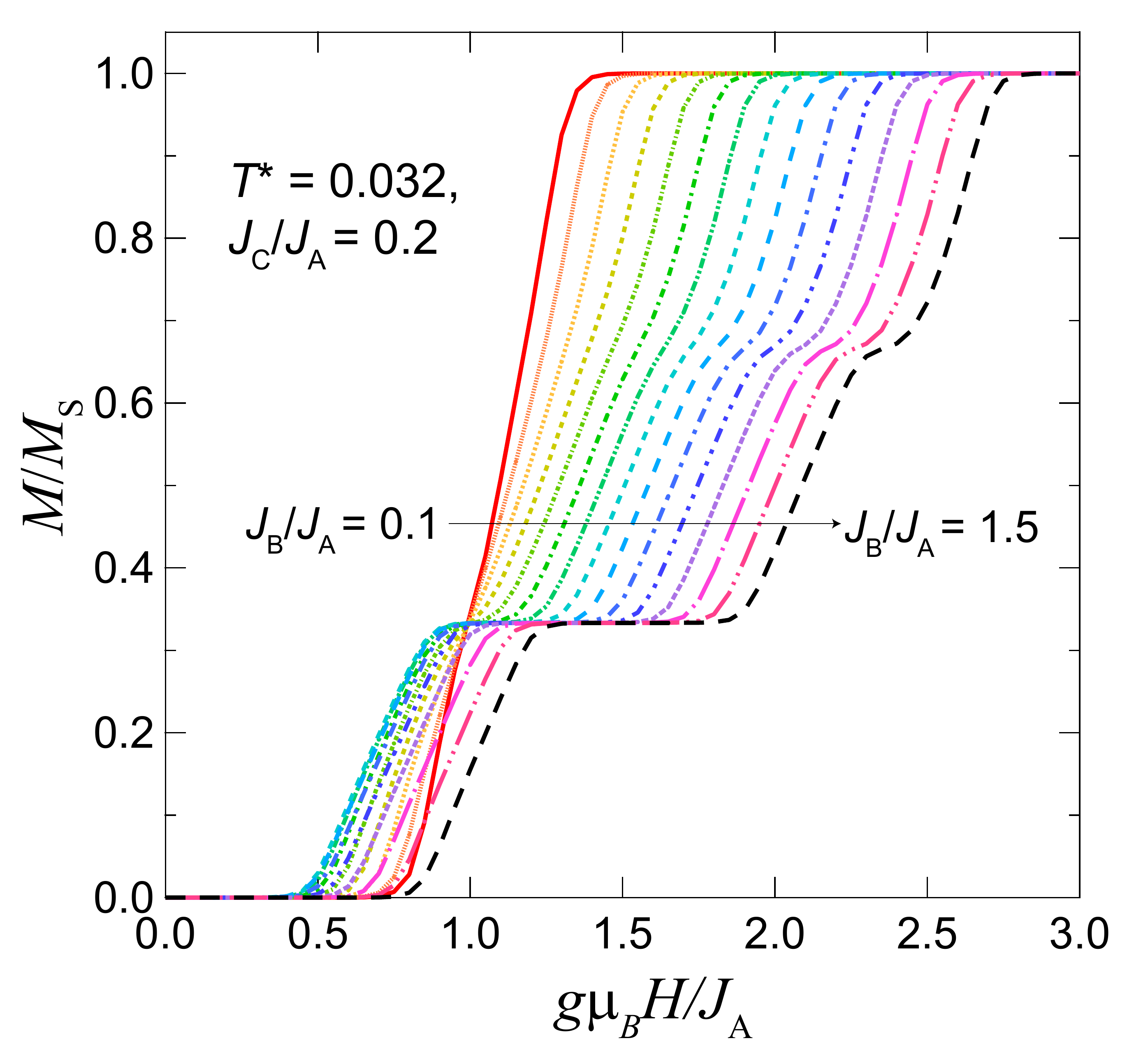}
\caption{Calculated magnetization curves for $J_{\rm B}/J_{\rm A}$ values (from 0.1 to 1.5 every 0.1 step) at a fixed value of $J_{\rm C}/J_{\rm A} = 0.2$. $M_{\rm S}$ represents the saturation magnetization and the magnetization $M$ is normalized by $M_{\rm S}$. The horizontal axis indicates the external magnetic field $H$ normalized by the exchange constant $J_{\rm A}$.}
\label{f5}
\end{figure}

Finally, let us discuss how to change the magnetization plateaus based on the QMC calculations. In our magnetization measurements, we observed one-third and two-thirds magnetization plateaus in Cu$_2$(pymca)$_3$(ClO$_4$). Therefore, we investigate the magnetization curves by changing the ratios of exchange bonds. Figure \ref{f5} shows the results of numerical calculations. In this figure, we fix $J_{\rm C}/J_{\rm A}=0.2$. The field range of each magnetization plateau increases with increasing the value of $J_{\rm B}/J_{\rm A}$. The observed magnetization curve is reproduced most when $J_{\rm C}/J_{\rm A} = 0.2$ and $J_{\rm B}/J_{\rm A} = 1$. This finding corresponds to the fact that two of three Cu-Cu exchange bonds have similar bond length as stated in the introduction. From the calculations, we have found that one small and two large exchange bonds are required to observe the $1/3$ and $2/3$ magnetization plateaus in a distorted HLA. In these calculations, another step-like magnetization above the $2/3$ magnetization plateau is not reproduced, and this may inspire further theoretical and experimental studies on this compound. Experimental studies in magnetic fields beyond 70~T would be required and more sophisticated model suitable for this compound needs to be considered. 


In conclusion, we have performed magnetic susceptibility, specific heat, ESR, and magnetization measurements of a polycrystalline sample of Cu$_2$(pymca)$_3$(ClO$_4$) which is regarded as the spin $1/2$ honeycomb-lattice antiferromagnet. Magnetic susceptibility shows a broad maximum at about 25~K, which is typical of a low dimensional antiferromagnet, and the specific heat at zero field smoothly decreases down to 0.6~K, indicating no long range magnetic order down to this temperature. The ESR spectra of the polycrystalline sample can be fitted with a single Lorentzian function, representing an exchange coupled isotropic antiferromagnet. High-field magnetization up to 70~T shows triple stepwise jumps and is considerably reproduced, except the magnetization near 70~T, by a Quantum Monte Carlo calculation using the following exchange constants $J_{\rm A}/k_{\rm B} = 43.7~{\rm K}$, $J_{\rm A} = J_{\rm B}$ and $J_{\rm C}/J_{\rm A} = 0.2$. In this study, we have found that the distortion of the honeycomb lattice, resulting in having two large and one small exchange interactions, causes the stepwise magnetization. 

\begin{acknowledgments}

AO and MH thank Prof. Sawa and Dr. Sugawara (present address: KEK) at Nagoya University for useful discussion on the structure of this compound. This work was partially supported by the Grant-in-Aid for Scientific Research (No. 24244059) and was carried out at the Center for Advanced High Magnetic Field Science in Osaka University under the Visiting Researcher's Program of the Institute for Solid State Physics, The University of Tokyo.

\end{acknowledgments}

\end{document}